\documentclass[sigconf]{acmart}
\usepackage{graphicx} %
\usepackage{hyperref}

\title[Mind \& Motion Workshop]{Mind \& Motion: Opportunities and Applications of Integrating Biomechanics and Cognitive Models in HCI}

\author{Arthur Fleig}
\authornote{Equal contribution}
\orcid{0000-0003-4987-7308}
\affiliation{
 \institution{ScaDS.AI, Leipzig University} 
 \city{Leipzig}
 \country{Germany}
}
\email{arthur.fleig@uni-leipzig.de}

\author{Florian Fischer}
\authornotemark[1]
\orcid{0000-0001-7530-6838}
\affiliation{
 \department{Department of Engineering}
 \institution{University of Cambridge}
 \city{Cambridge}
 \country{United Kingdom}
 }
\email{fjf33@cam.ac.uk}

\author{Markus Klar}
\authornotemark[1]
\orcid{0000-0003-2445-152X}
\affiliation{
 \department{School of Computing Science}
 \institution{University of Glasgow}
 \city{Glasgow}
 \country{United Kingdom}
 }
\email{markus.klar@glasgow.ac.uk}

\author{Patrick Ebel}
\orcid{0000-0002-4437-2821}
\affiliation{
 \institution{ScaDS.AI, Leipzig University} 
 \city{Leipzig}
 \country{Germany}
 }
\email{ebel@uni-leipzig.de}
 
\author{Miroslav Bachinski}
\orcid{0000-0002-2245-3700}
\affiliation{
 \department{Department of Information Science and Media Studies}
 \institution{University of Bergen}
 \city{Bergen}
 \country{Norway}
 }
\email{miroslav.bachinski@uib.no}

\author{Per Ola Kristensson}
\orcid{0000-0002-7139-871X}
\affiliation{
 \department{Department of Engineering}
 \institution{University of Cambridge}
 \city{Cambridge}
 \country{United Kingdom}
 }
\email{pok21@cam.ac.uk}

\author{Roderick Murray-Smith}
\orcid{0000-0003-4228-7962}
\affiliation{
  \department{School of Computing Science}
  \institution{University of Glasgow}
  \city{Glasgow}
  \country{United Kingdom}
  }
\email{Roderick.Murray-Smith@glasgow.ac.uk}

\author{Antti Oulasvirta}
\orcid{0000-0002-2498-7837}
\affiliation{
  \institution{Aalto University}
  \city{Helsinki}
  \country{Finland}
  }
\email{antti.oulasvirta@aalto.fi}
\renewcommand{\shortauthors}{Arthur Fleig et al.}
\copyrightyear{2025}
\acmYear{2025}
\setcopyright{rightsretained}
\acmConference[UIST Adjunct '25]{The 38th Annual ACM Symposium on User Interface Software and Technology}{September 28-October 1, 2025}{Busan, Republic of Korea}
\acmBooktitle{The 38th Annual ACM Symposium on User Interface Software and Technology (UIST Adjunct '25), September 28-October 1, 2025, Busan, Republic of Korea}\acmDOI{10.1145/3746058.3758473}
\acmISBN{979-8-4007-2036-9/2025/09}

\begin{document}

\begin{abstract}

Computational models of how users perceive and act within a virtual or physical environment offer enormous potential for the understanding and design of user interactions. %
Cognition models have been used to understand the role of attention and individual preferences and beliefs on human decision making during interaction, while biomechanical simulations have been successfully applied to analyse and predict physical effort, fatigue, and discomfort. 
The next frontier in HCI lies in connecting these models %
to enable robust, diverse, and representative simulations of different user groups.
These embodied user simulations could predict user intents, strategies, and movements during interaction more accurately, benchmark interfaces and interaction techniques in terms of performance and ergonomics, and guide adaptive system design.
This UIST workshop explores ideas for integrating computational models into HCI and discusses use cases such as UI/UX design, automated system testing, and personalised adaptive interfaces. 
It brings researchers from relevant disciplines together to identify key opportunities and challenges as well as feasible next steps for bridging mind and motion to simulate interactive user behaviour. %

\end{abstract}

\maketitle

\section{Motivation}
As the pace of technological development accelerates and user preferences become more diverse, there is a growing need in the Human-Computer Interaction (HCI) community to better understand how users' cognitive processes and physical actions \textit{interact} to shape user experience \cite{MurraySmith22, oulasvirta2022computational}. 
Cognitive models -- addressing mental processes like attention, memory, decision-making, and perception~\cite{lake2015human, kording2007causal, griffiths2010probabilistic, lewis2014computational, Lorenz2024} -- and biomechanical models -- focusing on motor actions, ergonomics, and physical constraints~\cite{Saul14, Holzbaur05, Fischer21, Caggiano22, lee19, schumacher2023:deprl} -- have made substantial contributions to user-centred design \cite{chen2015emergence, montano2017erg, evangelista2021xrgonomics, li2024nicer, cheema20, li2023modeling, fischer2024sim2vr}, yet they have largely evolved in isolation.

In the past, \textbf{biomechanical simulations} within HCI concentrated on inverse approaches to simulate kinematics, dynamics or muscle control based on recorded motion-capture data, thereby enabling the forecasting of fatigue or muscle utilisation \cite{Bachynskyi14, honglun2007research, bachynskyi2015performance}. 
More recently, biomechanical forward simulations have been employed, providing insights into human interaction movements \emph{prior} to conducting user studies~\cite{Ikkala22, cheema20, Fischer21, Klar23, 10.1145/3706599.3719699,10.1145/3746059.3747779}.
Advances in biomechanics (e.g., MyoSuite)~\cite{Caggiano22, schumacher2023:deprl, miazga2025increasing} open up new possibilities for generating more precise and more human-like interaction movements.

Concurrently, the field of \textbf{computational models of human cognition} significantly advanced, enabling predictions on the manner in which humans move \textit{and} on the underlying cognitive processes that drive these interactions. %
\textit{Computational Rationality} explores how humans make boundedly rational decisions by weighing costs and benefits in interaction~\cite{lewis2014computational, oulasvirta2022computational}. \textit{Active Inference}~\cite{Friston2016AIF,Parr2022AIF,stein2014AIF,murraysmith2024activeinferencehumancomputerinteraction} provides a framework for understanding how users learn, make predictions, and adjust their behaviour based on minimising prediction errors during interaction and Bayesian inference.

In light of these developments, the HCI community is on the cusp of achieving user models that integrate sophisticated models of biomechanics, perception, and cognition~\cite{Ikkala22, moon2023amortized}. 
An important missing step is finding a consensus within the HCI community on %
how to benchmark and evaluate user simulations, which poses a critical challenge for advancing their use.
At the same time, the potentially large differences in what these simulations should yield in particular use cases (e.g., in terms of scope and accuracy) make it challenging to find a one-size-fits-all benchmark. 

This workshop aims to generate actionable insights into how cognitive and biomechanical user models can solve HCI problems, improve real-world interface design, and be brought to the forefront of HCI practice. 
By bringing together researchers from computational interaction, UI/UX design, adaptive systems, biomechanics, and others, we will encourage discussions about the benefits and challenges of computational user models for understanding, predicting, and systematically analysing interactive behaviour.

We want participants to leave with practical knowledge, a clearer understanding of real-world applications that yield the most "bang for the buck" 
from %
biomechanical and cognitive models, and a roadmap for integrating these insights into research and design.

\section{Workshop Goals and Activities}

The primary goal of \textit{Mind \& Motion} is to deepen the community’s understanding of how biomechanical and cognitive models can be applied to enhance user interface design and usability. 
In addition to facilitating knowledge-sharing and inspiring new collaborations across disciplines, this workshop aims to identify practical applications where behavioural and embodied user models can drive significant improvements in HCI.
Our call\footnote{\url{https://mind-and-motion.github.io}} welcomes all researchers and practitioners interested in simulations and predictive models of interactive user behaviour. 
We particularly encourage participants to showcase hands-on examples of their recent work in these areas -- successful or not --, to give an idea of the ongoing activities within and beyond the HCI community.

\subsection{Planned Activities}

The workshop includes two keynotes %
on cutting-edge research in biomechanical simulations and cognitive modelling. %
After the keynotes, which intentionally follow each other to provide input from different areas and perspectives, we foster discussions and exchange in small groups.
This will focus on the two main objectives of this workshop: (i) to identify the most promising applications and use cases for biomechanical and cognitive user models \textit{(Brainstorming)}, and (ii) to discuss the main barriers and challenges that need to be addressed for successfully integrating model-based user simulations into prevailing design methods and frameworks \textit{(Group Discussion)}.
The results of these activities will be presented and debated in plenary.
We expect 25-40 participants and will form small breakout groups to ensure that everyone actively contributes during brainstorming, demo carousel, and discussion sessions.

\paragraph{Opening and Introductions (1 hour).}
The workshop organisers welcome the participants, briefly introduce themselves, and provide an overview of the workshop's agenda. 
In lightning presentations (1-2 min), each attendee will briefly introduce themselves, with a particular focus on their interest in and expectations of the workshop. 
This familiarises participants and sets common expectations. %

\paragraph{Keynote 1: Cognitive Modelling (45 min):}
This keynote will give insights into existing cognitive theories and models that already are or could be used in HCI. 
It will be held by an expert in cognitive science, with a particular focus on HCI.
The talk should take 30 min.\ to leave room for questions.
Reflecting on questions from facilitators such as 'How might these decision-making models affect physical user effort?' then tie cognition to physical action (Keynote 2).

\paragraph{Keynote 2: Biomechanical Simulations (45 min):} This keynote will be held by an expert in the field of human biomechanical simulations %
outside of the HCI community (for which we have already secured funding to cover the conference and workshop fees), such as Vittorio Caggiano, a senior researcher of MyoSuite\footnote{\href{https://sites.google.com/view/myosuite}{https://sites.google.com/view/myosuite}}.
The aim is to demonstrate the current state-of-the-art and to inspire participants to explore these simulations in the field of HCI.
Thus, roughly half of the 30 min.\ talk should use a mini live tool walk-through or recorded showcase, with room for interaction and discussion. 

\paragraph{Brainstorming: Use Cases and Applications (1 hour):} 
With both keynotes in mind, this session provides an opportunity to brainstorm ideas in smaller groups. 
It comprises two phases: a phase of collaborative work and a phase of presentation.
In the first phase, groups should discuss possible applications of biomechanical and cognitive models. 
We encourage to explore ideas across areas like UI/UX design, automated system testing, or adaptive interfaces, but groups may expand to other contexts too, with guiding questions such as 'If you had a perfect simulation of a human user, what are the most promising use cases? What does it require for the simulation to be applied here?'.
In the second phase, the group should summarise their results and briefly share them with the plenum, e.g., by sketching a simple example workflow or prototype that shows how the simulation could be used in a real HCI scenario.

\paragraph{Spotlight and Demo Session (1 hour)} 

In our application questionnaire, we will ask whether participants want to
share concrete examples of how they have used, tested, or prototyped cognitive and biomechanical models, including ideas that did not fully succeed. 
In this after-lunch session, instead of traditional short talks, selected presenters will showcase their work in a rotating "demo carousel". 
Small groups will move between stations, engaging with 4-6 mini-demos or tool showcases. 
We specifically invite presenters to bring code snippets, prototypes, datasets, or failed experiments to kick off discussions about ongoing challenges and common pitfalls. %

\paragraph{Group Discussion: Challenges and Limitations (1 hour):} This provides a chance to discuss about the grand challenges that simulations will face in the previously identified applications, such as the necessity for better models, computational resources, etc. 
Participants regroup and 
we will foster discussions %
by asking questions such as 'What are possible pitfalls or limitations for user simulations? What prevents you from using simulated users?'.
The intended output is a \textit{Challenge Poster} per group, outlining technical gaps, modelling needs, or open research questions, and developing initial ideas for minimal evaluation standards.

\paragraph{Closing Session (1 hour):} %
To round up the workshop and close on a reflective and action-oriented note, we will begin by clustering the identified use cases, challenges, and potential evaluation directions. 
We will then invite participants to individually share the idea or use case they found most compelling, and one tangible next step, such as testing a new model they have learnt about. 
We will conclude by summarizing shared interests and open the floor for brief discussion on how to sustain momentum, e.g., forming working groups. %

The day concludes with an optional informal dinner to continue conversations and build connections beyond the workshop.

\section{Post-Workshop Plans}
We will compile the results, such as challenge posters and workflow sketches, into a shared document accessible to all participants. %
Moreover, we plan to disseminate the knowledge and insights gained to the wider community in the form of an ACM Interactions article and through social media. 
We will invite all participants to join our Slack channel to stay connected. %
Depending on the discussion outcomes, we may also set up a shared repository (e.g., on GitHub) where attendees can upload cognitive and biomechanical models, evaluation tools, and datasets discussed during the workshop.

\bibliography{bib}

\appendix

\end{document}